\documentclass[journal]{IEEEtran}
\usepackage{graphicx, times, amsmath, amssymb, cite, subfigure, stfloats, subfigure,booktabs, mathtools,bm}
\usepackage[ruled,norelsize]{algorithm2e}

\makeatletter
\newcommand{\removelatexerror}{\let\@latex@error\@gobble}
\makeatother

\begin{document}

\title{EEG-Based Driver Drowsiness Estimation\\ Using Feature Weighted Episodic Training}

\author{
    Yuqi~Cui, Yifan~Xu and  Dongrui~Wu
    \thanks{
        Y.~Cui, Y.~Xu and D.~Wu are with the Key Laboratory of the Ministry of Education for Image Processing and Intelligent Control, School of Artificial Intelligence and Automation, Huazhong University of Science and Technology, Wuhan 430074, China. Email: yqcui@hust.edu.cn, yfxu@hust.edu.cn, drwu@hust.edu.cn.}
    \thanks{
    	Yuqi~Cui and Yifan~Xu contributed equally to this work.
        Dongrui~Wu is the corresponding author.
    }
}

\maketitle

\begin{abstract}
Drowsy driving is pervasive, and also a major cause of traffic accidents. Estimating a driver's drowsiness level by monitoring the electroencephalogram (EEG) signal and taking preventative actions accordingly may improve driving safety. However, individual differences among different drivers make this task very challenging. A calibration session is usually required to collect some subject-specific data and tune the model parameters before applying it to a new subject, which is very inconvenient and not user-friendly. Many approaches have been proposed to reduce the calibration effort, but few can completely eliminate it. This paper proposes a novel approach, feature weighted episodic training (FWET), to completely eliminate the calibration requirement. It integrates two techniques: feature weighting to learn the importance of different features, and episodic training for domain generalization. Experiments on EEG-based driver drowsiness estimation demonstrated that both feature weighting and episodic training are effective, and their integration can further improve the generalization performance. FWET does not need any labelled or unlabelled calibration data from the new subject, and hence could be very useful in plug-and-play brain-computer interfaces.
\end{abstract}


\section{Introduction} \label{sec:intro}

Driving safety is very important to our everyday life. However, according to the World Health Organization\footnote{https://www.who.int/violence\_injury\_prevention/road\_safety\_status/2018\newline/English-Summary-GSRRS2018.pdf} ``Global Status Report on Road Safety 2018", ``\emph{the number of road traffic deaths continues to rise steadily, reaching 1.35 million in 2016. ... Road traffic injuries are the eighth leading cause of death for all age groups. More people now die as a result of road traffic injuries than from HIV/AIDS, tuberculosis or diarrhoeal diseases. Road traffic injuries are currently the leading cause of death for children and young adults aged 5--29 years.}"

In addition to the reliability of the vehicle and the driver's experience, driving safety is also strongly related to the driver's alertness (or, drowsiness). Drowsy driving is the fourth major contributor to road crashes, following only to alcohol, speeding, and inattention \cite{sagberg2004fatigue}. Drowsiness impacts the driver's ability to quickly and appropriately respond to road emergencies, and hence may lead to accidents \cite{kozak2006evaluation}. Therefore, accurate estimation of the driver's drowsiness level is very important in preventing road accidents.

Many approaches have been reported~\cite{abbood2014prediction, chacon2015detecting,kang2013various,sahayadhas2012detecting,wang2014developing}, which can be roughly categorized into two directions: contactless detections and wearable sensor based detections. The former use cameras and/or other sensors, which are not attached to the driver's body, to  monitor the driver's facial activities and/or driving patterns to estimate the drowsiness level~\cite{bergasa2006real, d2007visual, sahayadhas2012detecting}. The latter use wearable sensors to measure the driver's physiological signals, e.g., electroencephalogram (EEG)~\cite{michail2008eeg}, electrocardiography (ECG)~\cite{jahn2005peripheral, michail2008eeg}, electromyography (EMG)~\cite{akin2008estimating, hu2009driver}, etc, and then perform drowsiness estimation.  The heart rate and heart rate variability can be easily obtained from ECG signals. They both vary significantly between alertness and drowsiness, and hence can be indicators of drowsiness \cite{fujiwara2019heart,miyaji2009driver}. EMG signal is usually combined with other signals to determine the drowsiness level. For instance, Lee~\emph{et al.}~\cite{boon2015mobile} proposed a driver fatigue detection approach using EMG and galvanic skin responses. Fu~\emph{et al.}~\cite{fu2016dynamic} proposed to use EEG, EMG and respiration signals to dynamically detect driver fatigue. In this paper, we focus on using only EEG signals for driver drowsiness estimation.

Since EEG directly measures the brain states, it is very suitable for human psychophysiological state evaluation \cite{arico2017passive}. The power spectrum of EEG has been used to estimate driver drowsiness level \cite{cui2017eeg,chuang2018dynamically,wu2015online,drwuTFS2017}, especially the theta (4-7Hz) and alpha (8-12Hz) bands \cite{chuang2018brain, klimesch2007eeg,arico2017passive}. Additionally, different brain regions have different abilities in assessing the driver's drowsiness level. Previous studies have shown that theta and alpha band activities in the central and occipital regions are more correlated to fatigue~\cite{perrier2016driving, zhao2012electroencephalogram, lin2012neurocognitive}. These results indicate that it may be beneficial to give different brain regions different weights in drowsiness estimation.

A major challenge in EEG-based driver drowsiness estimation is that, due to individual differences, it is very difficult to develop a generic estimator, whose parameters are fixed and optimal for all subjects. Hence, a subject-specific calibration session is usually required to tune the estimator, which is time-consuming and not user-friendly. Lots of efforts have been made to reduce or eliminate this calibration. One of the most frequently used approach is transfer learning~\cite{pan2010survey, azab2018review}, which uses data from other subjects/sessions (called source domains) to facilitate the learning for a new subject (called target domain). For instance, Lin and Jung \cite{lin2017improving} proposed a conditional transfer learning framework to promote positive transfer for each individual. It first assesses an individual's transferability for positive transfer, and then selectively leverages the data from others with comparable feature spaces. This approach has demonstrated promising performance in EEG-based emotion classification. Zanini \emph{et al.} \cite{zanini2018transfer} proposed a Riemannian space transfer learning framework, which uses a reference covariance matrix at the resting state to align data from different domains, before applying a Riemannian space classifier. He and Wu \cite{he2018transfer} proposed a similar EEG data alignment approach in the Euclidean space, which is more efficient than the Riemannian space data alignment approach, and can be used as a pre-processing step before any Euclidean space classifier. However, all these approaches considered only classification problems, and all required some labeled or unlabeled data from the target subject for calibration. So, they cannot be used in true plug-and-play brain-computer interfaces.

This paper considers a much more realistic, also more challenging, scenario: there are no calibration data (either labeled or unlabeled) from the target subject at all; we want to build a model from the auxiliary subjects and apply it directly to the target subject. Each auxiliary subject can be viewed as an independent source domain, and this problem setting is called \emph{domain generalization} in computer vision.

Many neural network based approaches have been proposed in recent years for domain generalization~\cite{ghifary2015domain, li2018domain,li2017deeper,li2019episodic,li2018learning,balaji2018metareg}, which can be summarized into two categories:
\begin{enumerate}
\item Train a robust cross-domain model using a specially designed neural network architecture to reduce the domain shift. For instance, Ghifary \emph{et al.} \cite{ghifary2015domain} proposed multi-task auto-encoder, which learns to transform the image in one domain into analogs in multiple related domains. These features, which are robust to variations across domains, are then fed into a classifier. Li \emph{et al.} \cite{li2017deeper} proposed a low-rank parameterized convolutional neural network to compensate the domain shift. Li \emph{et al.} \cite{li2018domain} used adversarial auto-encoders to align the distributions among different domains by minimizing the maximum mean discrepancy (MMD), and matched the aligned distribution to an arbitrary prior distribution via adversarial feature learning. The first step ensures the learned feature representation is universal to the known source domains, and the second step ensures the features can generalize well to the unseen target domain.

\item Train models with regularization or meta-learning scheme regardless of the model structure. Balaji \emph{et al.} \cite{balaji2018metareg} proposed a meta-regularization approach for domain generalization, which encodes domain generalization using a novel regularization function that makes the model trained in one domain to perform well in another domain. The regularization function was found in a learning-to-learn (or meta-learning) framework. Li \emph{et al.} \cite{li2018learning} proposed a model agnostic training procedure for domain generalization. Their algorithm simulated the shift between source and target domains during training by synthesizing virtual target domains within each mini-batch. The meta-optimization objective ensures performance improvements in both domains. Li \emph{et al.} \cite{li2019episodic} further proposed an episodic training (ET) procedure that trains a single deep network while exposing it to the domain shift that characterises a novel domain at runtime. Specifically, it decomposes a deep network into two components: feature extractor and classifier, and then trains each component by simulating it interacting with a partner which may not be well tuned for the current domain.
\end{enumerate}

This paper extends ET from classification to regression, and applies it to EEG-based driver drowsiness estimation. Our main contributions are:
\begin{enumerate}
\item We propose a feature weighting (FW) scheme that automatically assigns each feature a weight, by taking different importance of different brain regions into consideration.
\item We extend ET in \cite{li2019episodic} from classification to regression, and simplify it so that the computational cost is reduced without sacrificing the generalization performance.
\item We integrate FW and ET into a single learning framework, feature weighted episodic training (FWET), to achieve better generalization performance than each individual module.
\end{enumerate}

The remainder of this paper is organized as follows: Section~\ref{sec:method} introduces our dataset, feature extraction method, and the proposed FWET approach. Section~\ref{sec:results} evaluates the performance of FWET in EEG-based driver drowsiness estimation. Section~\ref{sec:conclusion} draws conclusion.

\section{Feature Weighted Episodic Training (FWET)} \label{sec:method}

\begin{figure*}[htpb] 	\centering
\includegraphics[width=0.95\linewidth]{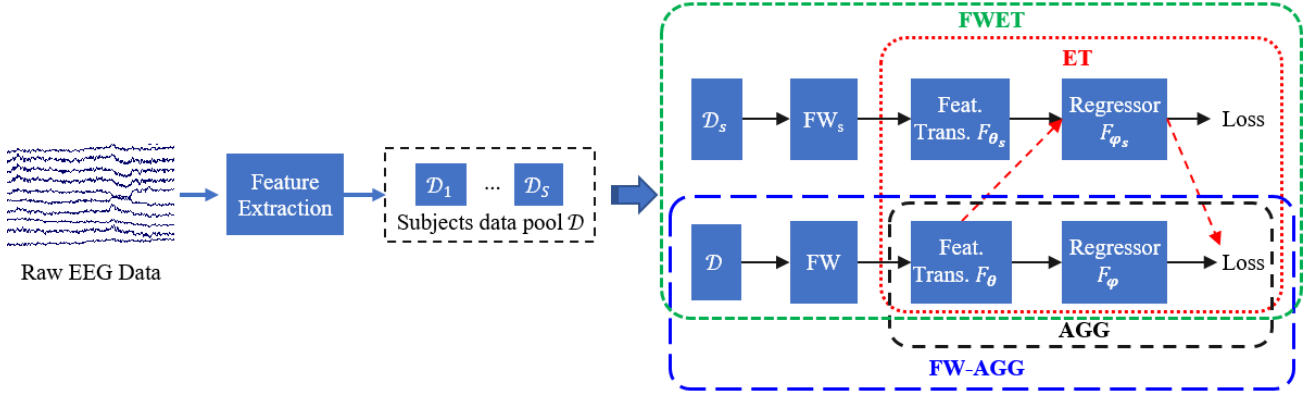}
\caption{AGG, FW-AGG, ET and FWET for EEG-based driver drowsiness estimation.}	\label{fig:algflow}
\end{figure*}

This section introduces the dataset for EEG-based driver drowsiness estimation, and our proposed FWET approach, whose overall flowchart is shown in Fig.~\ref{fig:algflow}, along with several other variants.

\subsection{Dataset}

\begin{figure}[htpb] \centering
\includegraphics[width=\linewidth]{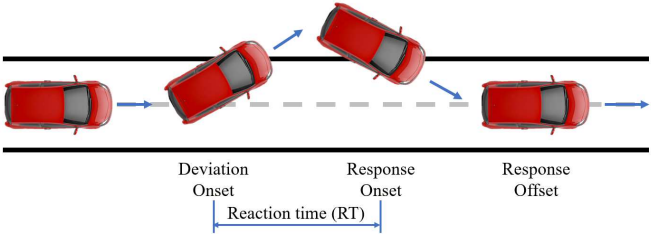}
\caption{Illustration of the way the reaction time was computed.} \label{fig:sketch}
\end{figure}

The data were collected in a simulated driving experiment, which was identical to that used in~\cite{wu2015online,cui2017eeg,chuang2014kinesthesia,chuang2012co}. Sixteen healthy subjects (age 24.2 $\pm$ 3.7, ten males, six females) with normal or corrected to normal vision were recruited to participate in a sustained-attention driving experiment \cite{chuang2014kinesthesia,chuang2012co}, which consisted of a real vehicle mounted on a motion platform with six degrees of freedom immersed in a 360-degree virtual reality scene. The experiment simulated driving on an empty highway at 100km/h. Every 5-10 seconds, a random lane-departure event was activated, which caused the car to drift from the center of the lane. The participants were asked to steer the car back to the center of the lane immediately. The reaction time was calculated as the time difference between the drift and the moment the subject started to act, as shown in Fig.~\ref{fig:sketch}. If the participant did not respond to the lane-departure event, such as falling asleep, the vehicle would hit the boundary of the road and continue moving forward along the boundary. The next lane-departure event happened after the response offset. Each participant performed the experiment for 60-90 minutes in the afternoon when the circadian rhythm of sleepiness reached its peak~\cite{horne1999vehicle}.

The Institutional Review Board of the Taipei Veterans General Hospital approved the experimental protocol. Each participant read and signed an informed consent form before the experiment began.

The reaction time $\tau$ was later converted into a drowsiness index (DI) \cite{wu2015online,drwuTFS2017,wei2015selective,Wei2016,Wei2018},
\begin{align}
DI = \max\left(0, \frac{1-e^{-(\tau-\tau_0)}}{1+e^{-(\tau-\tau_0)}}\right), 	\label{eq:di}
\end{align}
where $\tau_0$ was set to 1 in our work. The DIs were then smoothed by a 90s moving-average window. (\ref{eq:di}) maps the reaction time to $[0,1]$ and overcomes its long-tail effect (very large reaction time was rare, but it did exist; such extreme values would significantly deteriorate the overall estimation). The fatigue level has been demonstrated to have a strong correlation with the reaction time \cite{ji2004real}. Since the DI is positively correlated with the reaction time, DI is also an indicator of the fatigue level.

Note that the value of $\tau_0$ could also be set individually for each subject. For instance,  in~\cite{wei2015selective}, $\tau_0$ was set to the 5 percentile value of the reaction time in each session. However, in a real-world online plug-and-play brain-computer-interface system, we do not have training data from the target subject, thus setting $\tau_0$ individually is not possible. Nevertheless, to demonstrate the robustness of our proposed approach, we also compare the performances using $\tau_0=1$ and individualized $\tau_0$ in Section~\ref{sec:adaptive}, which is possible in offline driver drowsiness estimation.

During the experiment, EEG signals were recorded using a 500Hz 32-channel Neuroscan system (30-channel EEGs plus 2-channel earlobes). Since data from one subject were not recorded correctly, we only used 15 subjects in our paper. To ensure a fair comparison, we used the first 3,600 seconds data from each subject.

\subsection{Preprocessing and Feature Extraction}

We used EEGLAB~\cite{delorme2004eeglab} for data preprocessing. We first performed 1-50Hz band-pass filtering to remove artifacts and noise, and then down-sampled the data from 500Hz to 250Hz and re-referenced them to the averaged earlobes.

We tried to predict the DI for each subject every 3 seconds, using 30-second EEG signal before each sample point. We computed the average power spectral density (PSD; their absolute values, instead of relative values, were used) in theta and alpha bands using Welch's method~\cite{welch1967use}, with Hamming window, 1024 points fast Fourier transform, and 50\% overlapping. The PSDs were then converted into dBs and used as our features. Each feature vector had $30\times2=60$ dimensions. All algorithms in our experiments used the same PSD features described above.

Each 30-second EEG signal may include brain activities, e.g., visual stimulus of the lane departure event and the wheel steering intention, and interferences from the wheel steering motor execution and other body movements. These brain activities and interferences are inevitably happening in real-world driving scenarios, and a good drowsiness estimation algorithm should be able to cope with them. Moreover, there are some other activities that are normal in realistic driving situations but were not considered in our experiments, e.g., the motor executions of acceleration and braking, talking, etc. These should be considered in the future improved experiment design.

\subsection{Problem Setting}

Assume Subject $s$ has $N_s$ labeled EEG trials $\mathcal{D}_s=\{\textbf{x}_i^s, y_i^s\}_{i=1}^{N_s}$, where $\textbf{x}_i^s\in \mathbb{R}^{d\times1}$ is a $d$-dimension feature vector extracted from the $i$-th EEG trial of Subject $s$, and $y_i^s$ is the corresponding DI. Assume also that we have $S$ subjects in our training set, and we want to predict the DI for $\mathcal{D}_t=\{\textbf{x}_i^t\}_{i=1}^{N_t}$ from an unseen target subject $t$.

Our model contains two components: $F_{\bm{\theta}}$, the feature transformation network, and $F_{\bm{\psi}}$, the regression network. Hence the prediction for $\textbf{x}_i^t$ is $\hat{y}_i^t=F_{\bm{\psi}}(F_{\bm{\theta}}(\textbf{x}_i^t))$.

\subsection{Aggregation Training (AGG)}

The simplest domain generalization approach is to combine all source subjects' data to train one single model, which is usually a very strong baseline. This method is called aggregation training (AGG) in \cite{li2019episodic}.

In this paper, we perform AGG using a multi-layer perceptron (MLP) neural network with one hidden layer and ReLU activation function. The loss function is:
\begin{align}
\mathcal{L}_{AGG}=\sum_{s=1}^{S}\sum_{i=1}^{N_s}\ell(y_i^s,~F_{\bm{\psi}}(F_{\bm{\theta}}(\textbf{x}_i^s))), \label{eq:agg}
\end{align}
where $\ell$ is the squared error in regression. The parameters $\bm{\psi}$ and $\bm{\theta}$ are learned through gradient descent optimization.

\subsection{Feature Weighting (FW)}\label{subsec:fr}

Previous studies \cite{perrier2016driving, zhao2012electroencephalogram, lin2012neurocognitive} have shown that EEG features (channel-wise PSD features in this paper) in different brain regions have different correlations to the drowsiness. Thus, we use the following FW scheme to assign different weights to different EEG channels:
\begin{align}
\hat{w}_l&={e^{w_l}}/{\sum_{j=1}^{d}e^{w_j}},\quad l=1,...,d \\
\hat{\textbf{x}}_i^s &= \hat{\mathbf{w}}\circ \textbf{x}_i^s,\quad s=1,...,S \label{eq:xjs}
\end{align}
where $\mathbf{w}=[w_1,...,w_d]^T\in \mathbb{R}^{d\times1}$ and $\hat{\mathbf{w}}=[\hat{w}_1,...,\hat{w}_d]^T\in \mathbb{R}^{d\times1}$ are the original and transformed weight vectors, respectively, and $\circ$ denotes element-wise product. We do not use the weight $\mathbf{w}$ directly in (\ref{eq:xjs}); instead, we use its $softmax$ version $\hat{\mathbf{w}}$, to make sure the weights are non-negative and sum up to 1.

\subsection{Episodic Training (ET)}

ET for domain generalization was recently proposed by Li \emph{et al.} \cite{li2019episodic} for image recognition. We simplify their algorithm and integrate it with FW. The original ET algorithm in \cite{li2019episodic} contains three regularization terms. In our work, we only adopt the first loss term (described as \emph{epif} in Section~\ref{sec:etres}) for simplicity and speed. As it will be shown in Section~\ref{sec:discussion}, our simplification greatly reduces the computational cost of the original ET, without sacrificing its generalization performance.

A common approach in transfer learning to learn domain-invariant features is to train a feature extractor $F_{\bm{\theta}}$ that makes the marginal distribution $P(F_{\bm{\theta}}(\textbf{x}^s))$ consistent for different source domains, $s=1,...,S$. However, since the DIs of different subjects vary due to individualized differences, i.e., the conditional distributions $P(y^s|F_{\bm{\theta}}(\textbf{x}^s))$ are different for different subjects, aligning the marginal distributions only may not lead to satisfactory generalization performance. ET considers the conditional distributions $P(y^s|F_{\bm{\theta}}(\textbf{x}^s))$ directly, and trains an $F_{\bm{\theta}}$ that aligns $P(y^s|F_{\bm{\theta}}(\textbf{x}^s))$ in all source domains, which usually generalizes well to the unseen target domain $\mathcal{D}_t$.

We first establish a subject-specific feature transformation (FT) model $F_{\bm{\theta}_{s}}$ and a subject-specific regression model $F_{\bm{\psi}_{s}}$ for each source subject to learn the domain-specific information. We also want to train an FT model $F_{\bm{\theta}}$ that makes the transformed features from Subject~$s$ still perform well when applied to a regressor $F_{\bm{\psi}_j}$ trained on Subject~$j$ ($j\neq s$). Hence, the following loss function is used:
\begin{align}
\mathcal{L}_{FT}^{s,j} = \sum_{i=1}^{N_s}\ell(y_i^s, ~\overline{F}_{\bm{\psi}_j}(F_{\bm{\theta}}(\textbf{x}_i^s))), \label{eq:et}
\end{align}
where $\overline{F}_{\bm{\psi}_j}$ means that $F_{\bm{\psi}_j}$ is not updated during back propagation.

The overall loss function of ET, when Subject $s$'s data are fed into Subject~$j$'s regressor, is:
\begin{align}
\mathcal{L}_{ET}^{s,j} = \mathcal{L}_{AGG}^s+\lambda\cdot \mathcal{L}_{FT}^{s,j}. \label{eq:L}
\end{align}
where
\begin{align}
\mathcal{L}_{AGG}^s=\sum_{i=1}^{N_s}\ell(y_i^s,~F_{\bm{\psi}}(F_{\bm{\theta}}(\textbf{x}_i^s))). \label{eq:aggs}
\end{align}
$\lambda=0.1$ was used in our experiments.

Note that since there is a (purposeful) mismatch between $F_{\bm{\psi}}$ and $F_{\bm{\theta}}$ in $\mathcal{L}_{FT}^{s,j}$, the gradient $\partial \mathcal{L}_{FT}^{s,j}/\partial F_{\bm{\theta}}$ may be unstable and sometimes have gradient explosion. Therefore, we clipped the gradient $\partial \mathcal{L}_{FT}^{s,j}/\partial F_{\bm{\theta}}$ to $[-10, 10]$.

\subsection{FWET}

Our proposed algorithm, FWET, which integrates FW and ET, is shown in Algorithm~\ref{algo:ep}. It learns $\mathbf{w}$ in FW and $\bm{\theta}$ and $\bm{\psi}$ in ET simultaneously through gradient descent optimization.

All $\bm{\theta}$, $\bm{\psi}$, $\bm{\theta}_s$ and $\bm{\psi}_s$, $s=1,...,S$, are uniformly initialized. Take $\bm{\theta}$ as an example. Let $M$ be the number of features in each layer. Then, each element of $\bm{\theta}$ is initialized as a uniformly distributed random variable in $[-\sqrt{1/M},\sqrt{1/M}]$.

\begin{algorithm}
\KwIn{Training subject data $\{\textbf{x}_i^s, y_i^s\}_{i=1}^{N_s}$, $s=1,...,S$\;
 \hspace*{10mm} ET weight $\lambda$\;
 \hspace*{10mm} Batch size $N$\;
 \hspace*{10mm} Learning rate $\alpha$.}
\KwOut{FWET model parameters $\mathbf{w}$, $\boldsymbol{\theta}$ and $\boldsymbol{\psi}$.}
 \For{$s=1:S$}{
 Initialize domain-specific FW vector $\mathbf{w}_s=\mathbf{1}$\;
  Randomly initialize domain-specific model parameters $\bm{\theta}_s$ and $\bm{\psi}_s$\;
 }
 Initialize $\mathbf{w}=\mathbf{1}$ in FWET\;
 Randomly initialize $\bm{\theta}$ and $\bm{\psi}$ in FWET\;

 \tcp{Warm up}
 \For{$s=1:S$}
 {Train the domain-specific model parameters $\mathbf{w}_s$, $\bm{\theta}_s$ and $\bm{\psi}_s$ for one epoch, using only data from Subject~$s$\;}

 \While{training}{
  \For{$s=1:S$}{
   Sample a batch $\{\textbf{x}_i^s, y_i^s\}_{i=1}^N$ from Subject $s$\;
   Compute $\hat{\textbf{x}}_i^s$ using (\ref{eq:xjs}), $i=1,...,N$\;
   Compute the sum of squared loss for the domain-specific model $\mathcal{L}_{DS}=\sum_{i=1}^{N}\ell(y_i,F_{\bm{\psi}_s}(F_{\bm{\theta}_s}(\hat{\mathbf{x}}_i^s)))$\;
   $\mathbf{w}_s = \mathbf{w}_s - \alpha \nabla_{\mathbf{w}_s}\mathcal{L}_{DS}$\;
   $\bm{\theta}_s = \bm{\theta}_s - \alpha \nabla_{\bm{\theta}_s}\mathcal{L}_{DS}$\;
   $\bm{\psi}_s = \bm{\psi}_s - \alpha \nabla_{\bm{\psi}_s}\mathcal{L}_{DS}$\;
  }
  \For{$s=1:S$}{
   \For{$j=1:S$ and $j\not=s$}{
    Sample a batch  $\{\textbf{x}_i^s, y_i^s\}_{i=1}^N$  from Subject $s$\;
    Compute the loss $\mathcal{L}_{ET}^{s,j}$ in (\ref{eq:L}) on the batch\;
        $\mathbf{w} = \mathbf{w} - \alpha \nabla_\mathbf{w} \mathcal{L}_{ET}^{s,j}$\;
    $\bm{\theta} = \bm{\theta} - \alpha \nabla_{\bm{\theta}} \mathcal{L}_{ET}^{s,j}$\;
    $\bm{\psi} = \bm{\psi} - \alpha \nabla_{\bm{\psi}} \mathcal{L}_{ET}^{s,j}$\;
   }
  }
 }
 \caption{Pseudocode of FWET.}  \label{algo:ep}
\end{algorithm}

\section{Experimental Results}\label{sec:results}

This section studies the performance of FWET in EEG-based driver drowsiness estimation.

\subsection{Evaluation Method and Performance Measures} \label{sect:Eva}

We used leave-one-subject-out cross-validation to validate the performance of our model. Since this was a regression problem, we used two metrics to evaluate the prediction results: root mean squared error (RMSE) and Pearson correlation coefficient (CC), which respectively measure the error and the correlation between the predicted DIs and the groundtruth DIs.

We compared six different algorithms:
\begin{itemize}
	\item \emph{kNN}, which was a $k$-nearest neighbors regressor with $k=5$. The prediction was the average of the five nearest neighbors.
	\item \emph{RR}, which was ridge regression with L2 regularization coefficient $\alpha=0.1$.
	\item \emph{AGG}, which was an MLP neural network with only one hidden layer, trained using the loss function in (\ref{eq:agg}). The number of hidden layer units was 40.
	\item \emph{FW-AGG}, which performed FW before AGG.
	\item \emph{ET}, which trained an AGG model and $S$ domain-specific models together using ET. Each such model has the same structure as AGG above, i.e., a 3-layer MLP. The first layer was treated as $F_{\bm{\theta}}$. The other two layers were treated as $F_{\bm{\psi}}$.
	\item \emph{FWET}, which performed FW before ET.
\end{itemize}

The first two algorithms are commonly used baselines for regression problems. The last four are AGG based. We compare them to analyze the individual contributions of FW and ET in FWET. The last four models were trained using mini-batch gradient descent with momentum, with batch size $32$, learning rate $0.001$, momentum $0.9$, and weight decay $0.00005$. We sampled 10\% data from each training subject as the validation set in early-stopping to reduce overfitting. The maximum number of training epochs was set to 500, and early-stopping patience was 10 epochs. One epoch means one $training$ iteration in Algorithm~\ref{algo:ep}. We repeated all algorithms five times and report the average performance.

\subsection{Experimental Results}\label{subsec:expres}

\begin{figure*}[htpb] \centering
\subfigure[]{\includegraphics[width=0.95\linewidth,clip]{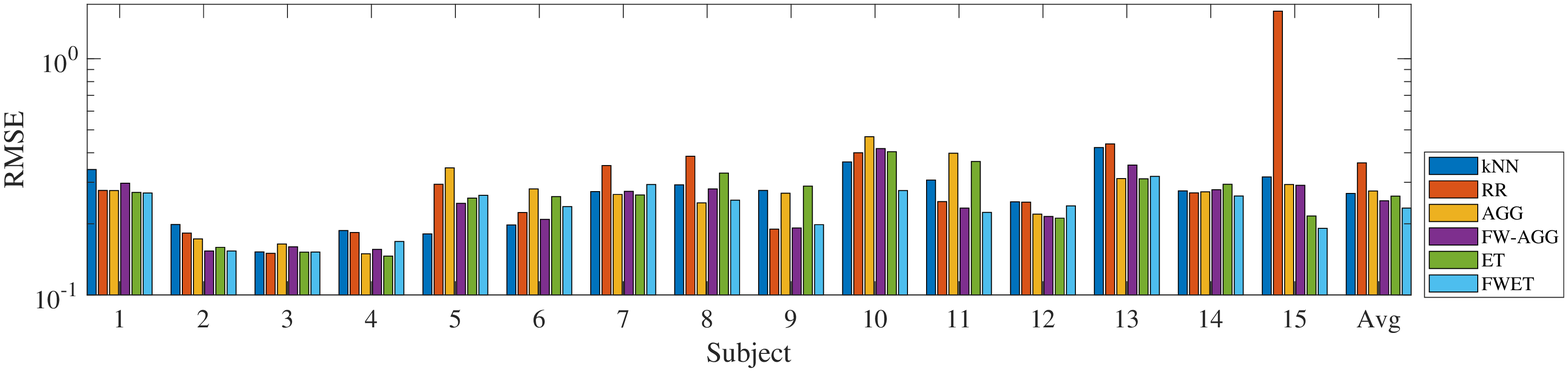}}
\subfigure[]{\includegraphics[width=0.95\linewidth,clip]{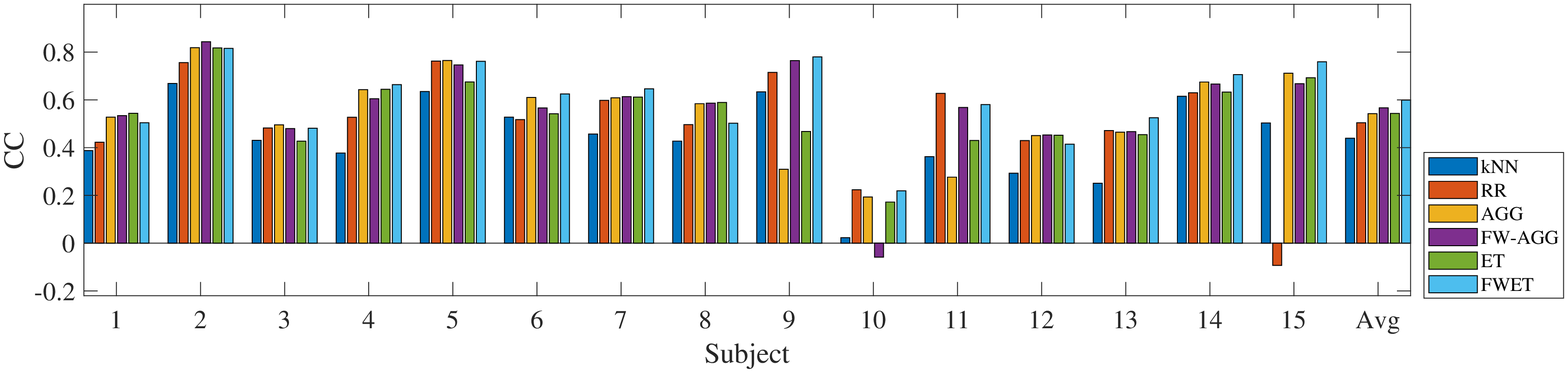}}
\caption{(a) RMSEs and (b) CCs in leave-one-subject-out cross-validation. The experiments were repeated five times, and the averages are shown.}	
\label{fig:bar_rmsecc}
\end{figure*}

The regression performance for each subject, averaged over five repeats, is shown in Fig.~\ref{fig:bar_rmsecc}. FW-AGG, ET and FWET outperformed kNN, RR and AGG for most subjects. One exception is Subject~10, on which FW-AGG and FWET gave negative CCs.

\begin{figure}[htpb] \centering
\subfigure[]{\includegraphics[width=0.4\linewidth,clip]{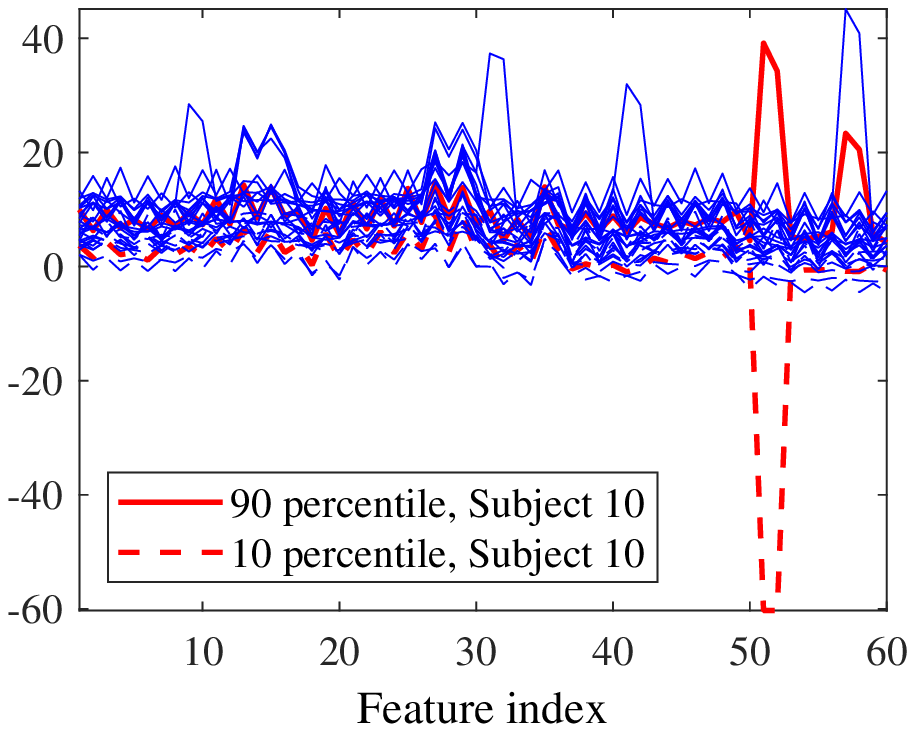}\label{fig:sub10a}}
\subfigure[]{\includegraphics[width=0.4\linewidth,clip]{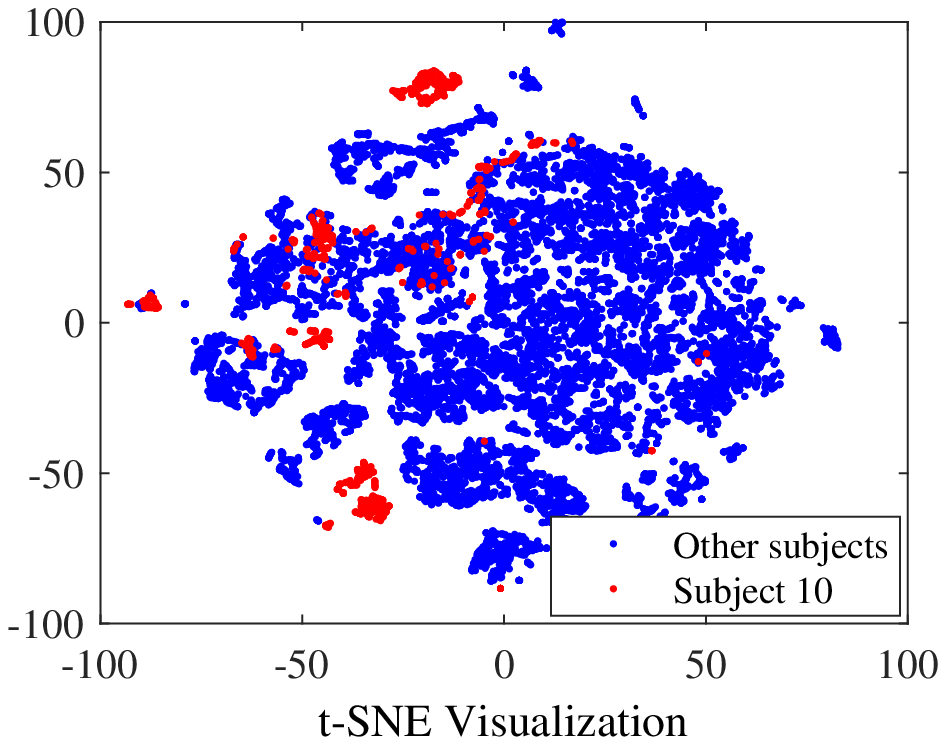}\label{fig:sub10b}}
\caption{(a) 90 and 10 percentiles of features from Subject~10, w.r.t. the corresponding feature percentiles (90: solid curves; 10: dashed curves) from the other 14 subjects; (b) $t$-SNE visualization of the features from different subjects.}	
\label{fig:sub10}
\end{figure}

To explore why FW-AGG and FWET gave weird CCs on Subject~10, we plot the feature distributions of Subject~10, along with those from the other 14 subjects, in Fig.~\ref{fig:sub10}. We first plot the 10 and 90 percentile of PSD features from each subject in Fig.~\ref{fig:sub10a} and  a $t$-SNE visualization in Fig.~\ref{fig:sub10b} to see if there are differences on feature distribution between subjects. Clearly, the distributions of the 51st and 52nd features of Subject~10 are dramatically different from those of other subjects, which may be due to outliers. We can also see that there are some data points from Subject~10 that are not consistent with the data from other subjects. The unsatisfactory performance of FW-AGG and FWET on Subject~10 suggests that maybe FW is sensitive to outliers.

\begin{figure}[htpb] \centering
\subfigure[]{\includegraphics[width=0.45\linewidth,clip]{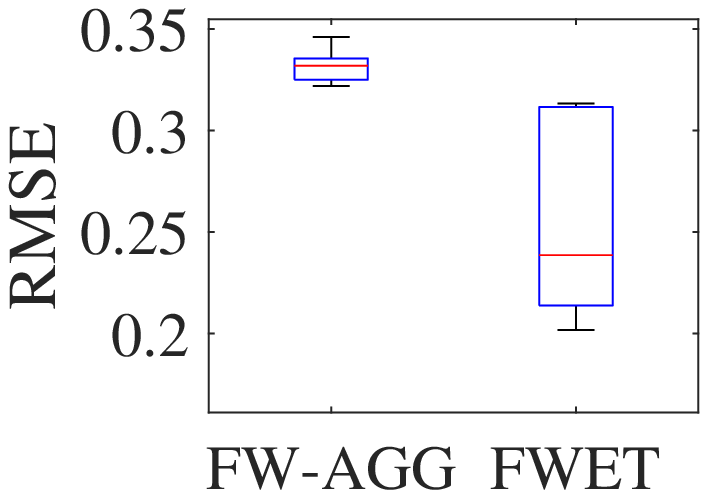}} \hfill
\subfigure[]{\includegraphics[width=0.45\linewidth,clip]{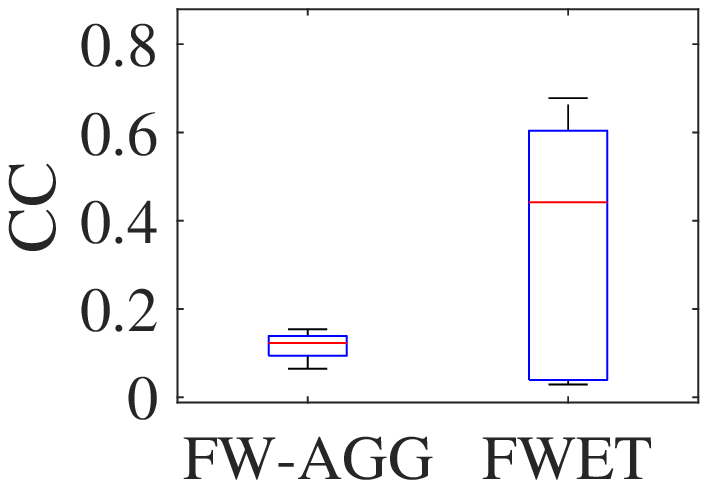}}
\caption{Boxplots of (a) RMSEs and (b) CCs of FW-AGG and FWET, when Subject~10 was the target (test) subject, and the 51st and 52nd outlier features were removed.}	 \label{fig:S10}
\end{figure}

In offline applications, we know the features from the target subject. So, preprocessing may be used to remove the outlier features. For example, when the 51st and 52nd outlier features of Subject~10 were removed, the corresponding boxplots of the RMSEs and CCs of FW-AGG and FWET are shown in Fig.~\ref{fig:S10}. They were considerably improved over the RMSEs and CCs in Fig.~\ref{fig:bar_rmsecc}.

The last group in each subfigure of Fig.~\ref{fig:bar_rmsecc} also shows the average performance across the 15 subjects, whose values are given in Table~\ref{tab:res_ave_rmsecc}. AGG is a nonlinear model with more parameters than kNN and RR. Theoretically, it should outperform kNN and RR if well-trained. However, Table~\ref{tab:res_ave_rmsecc} shows that this was not the case. AGG had slightly worse average RMSE than kNN, and slightly worse average CC than RR. There may be two reasons: 1) there were not enough training data to tune AGG well; and, 2) AGG was over-fitted on the training data, so it did not generalize well to a new subject. After introducing FW and ET, both training performance and generalization ability were improved, and both FW-AGG and FWET outperformed the three baselines (kNN, RR, and AGG). More specifically, ET outperformed AGG, improving 4.9\% and 0.2\% on the RMSE and the CC, respectively. After adding FW to AGG, FW-AGG further outperformed ET by 4.5\% on the RMSE and 4.3\% on the CC. FWET achieved the best performance, and further improved the RMSE and the CC by 6.9\% and 5.7\%, respectively, over FW-AGG.

\begin{table}[htpb] \centering
\caption{Average RMSEs and CCs across the 15 subjects and five runs.}
\begin{tabular}{c|ccccccc} \hline
      & kNN    & RR     & AGG    & FW-AGG & ET     & FWET \\      \hline
 RMSE & 0.2688 & 0.3622 & 0.2756 & 0.2504 & 0.2621 & \textbf{0.2332} \\
 CC   & 0.4394 & 0.5044 & 0.5422 & 0.5668 & 0.5434 & \textbf{0.5989} \\ \hline
\end{tabular} \label{tab:res_ave_rmsecc}
\end{table}

To determine if the differences between different algorithms were statistically significant, we also performed non-parametric multiple comparison tests on the RMSEs and CCs using Dunn's procedure \cite{Dunn1964}, with a $p$-value correction using the False Discovery Rate method \cite{Benjamini1995}. The results are shown in Table~\ref{tab:Dunn}, where the statistically significant ones are marked in bold. FWET statistically significantly outperformed kNN, RR and AGG on the RMSE, and also kNN and RR on the CC. Though the performance improvements of FWET over FW-AGG and ET were not statistically significant, we have seen from Fig.~\ref{fig:bar_rmsecc} and Table~\ref{tab:res_ave_rmsecc} that on average FWET still slightly outperformed them.

\begin{table}[htpb] \centering
\caption{$p$-values of non-parametric multiple comparisons on the RMSEs and CCs.}\label{tab:Dunn}
\begin{tabular}{c|l|ccccc}   \hline
&   & \texttt{kNN} &           \texttt{RR} &     \texttt{AGG}  & \texttt{FW-AGG} &  \texttt{ET}   \\ \hline
&\texttt{RR} & .3697 & & & &\\
&\texttt{AGG} & .4558 & .3814& & & \\
RMSE&\texttt{FW-AGG} & .1478 & .0936 & .1376& \\
&\texttt{ET} & .1982 & .1390 & .2146 &.3783 & \\
&\texttt{FWET} & \textbf{.0182} & \textbf{.0098} & \textbf{.0170} & .1745 &.1335 \\   \hline
&\texttt{RR} & \textbf{.0063} & & & &\\
&\texttt{AGG} & \textbf{.0000} & .0875 & & & \\
CC&\texttt{FW-AGG} & \textbf{.0000} & .0347 & .3043& \\
&\texttt{ET} & \textbf{.0001} & .1657 & .3440 &.1908 & \\
&\texttt{FWET} & \textbf{.0000} & \textbf{.0021} & .0873 & .1832 &.0403 \\   \hline
\end{tabular}
\end{table}

In summary, we have shown that it is always preferable to use FWET over the other five algorithms.

\subsection{Effects of FW}

\begin{figure}[htpb] \centering
\subfigure[]{\includegraphics[width=0.8\linewidth,clip]{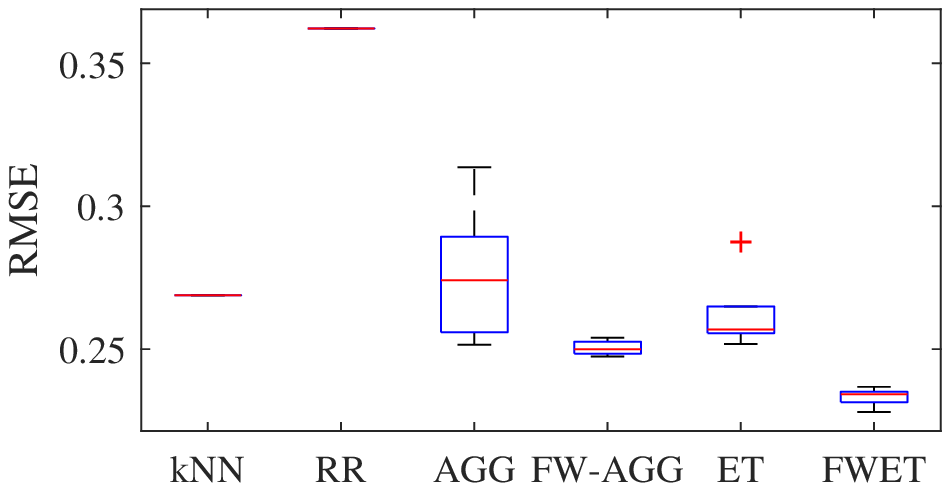}}
\subfigure[]{\includegraphics[width=0.8\linewidth,clip]{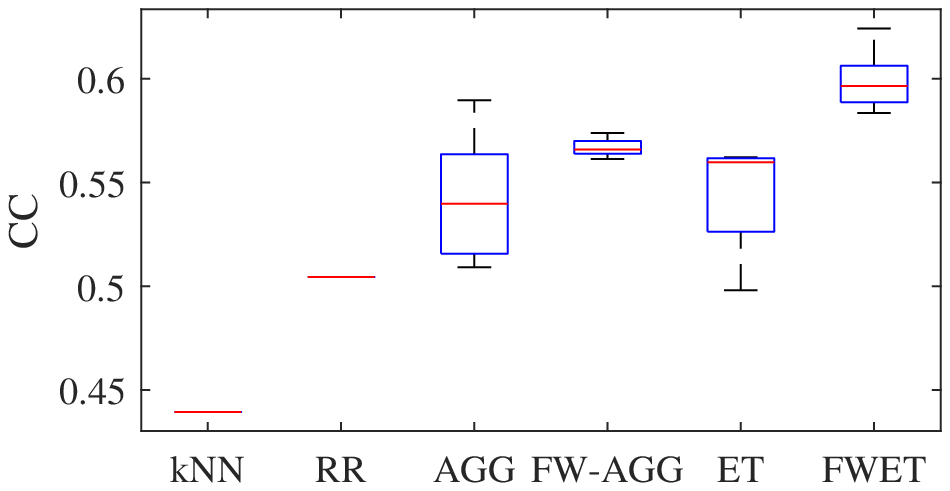}}
\caption{Boxplots of the average (a) RMSEs and (b) CCs of the six algorithms.}	
\label{fig:box_rmsecc}
\end{figure}

The four AGG based algorithms have randomness involved, e.g., initialization, batch selection, etc. It's interesting to study their stability. Recall that we had 15 subjects, and each algorithm was run five times when each subject was used as the target subject. The final results were assembled into a $15\times 5$ RMSE matrix and a $15\times 5$ CC matrix. We could plot a boxplot for each subject to show the stability of different algorithms, but that would take too much space, and is difficult to see the forest for the trees. So, we first computed the average performance of each algorithm over the 15 subjects, i.e., we took the average of the RMSE (CC) matrix along the columns to obtain a $1\times5$ row vector, and then plotted the box-plots of the five average RMSEs (CCs) in Fig.~\ref{fig:box_rmsecc}. The RMSEs and CCs of kNN and RR did not have uncertainty, because there was no randomness in these algorithms. Among the four AGG based algorithms, AGG and ET had large variations, and FW-AGG and FWET had very small variations, suggesting one more advantage of introducing FW to AGG, beyond better RMSE and CC.

Fig.~\ref{fig:box_rmsecc} shows that generally FW helped reduce the variation from different runs. It's interesting to study why. Several studies had analyzed the relationship between the generalization performance and sharp minima~\cite{chaudhari2017entropy,keskar2017large}. It is believed that sharp minima may lead to bad generalization performance. ET tends to have flatter minima, which had already been demonstrated in \cite{li2019episodic}. We want to investigate if FW has a similar effect. We added random Gaussian noise to the learned parameters and checked how quickly the performance degraded. A rapid decrease indicates that the model is at a sharp minimum, which is bad for generalization.

\begin{figure}[htpb]	\centering
	\subfigure[]{\includegraphics[width=0.48\columnwidth,clip]{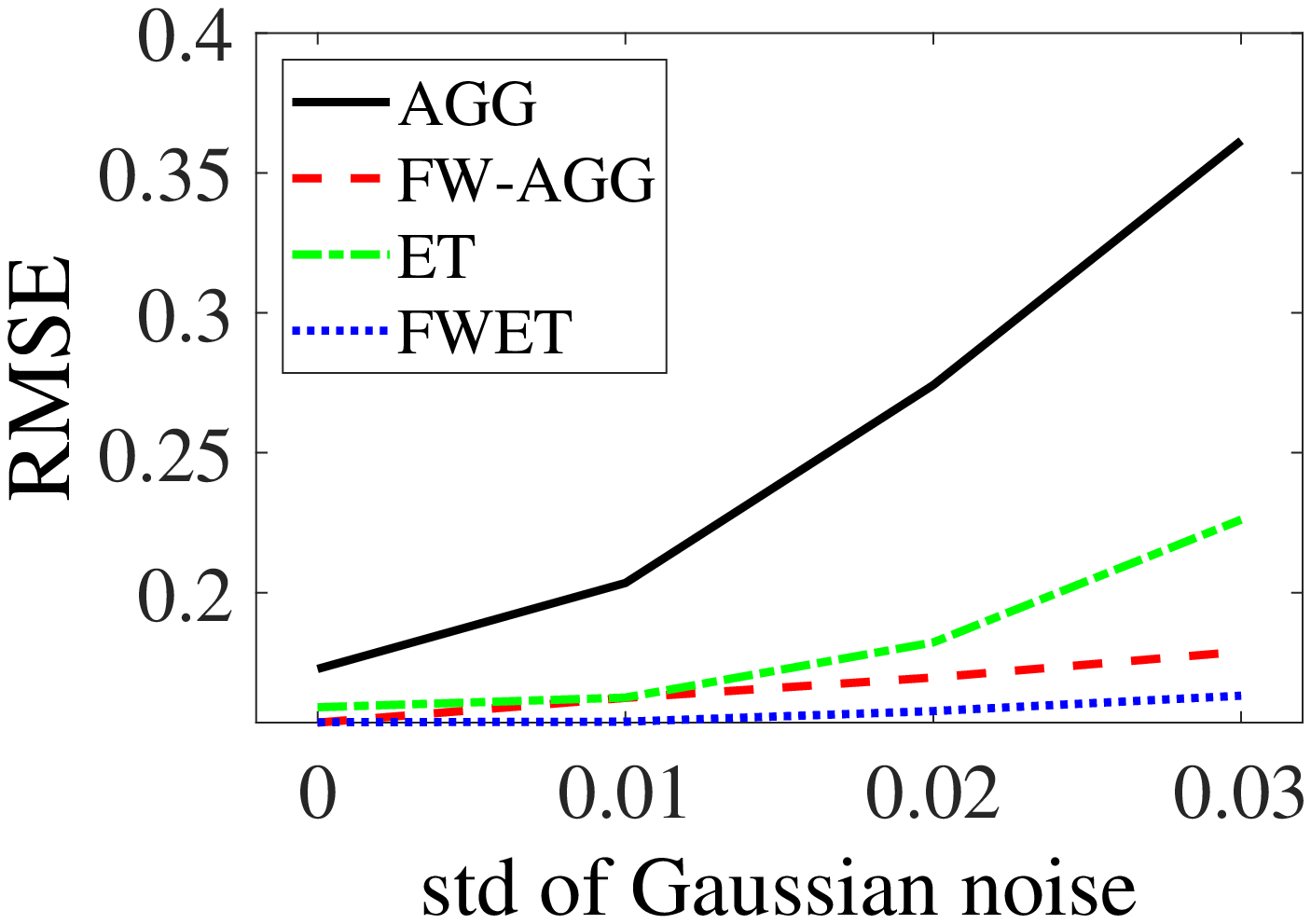}}
	\subfigure[]{\includegraphics[width=0.48\columnwidth,clip]{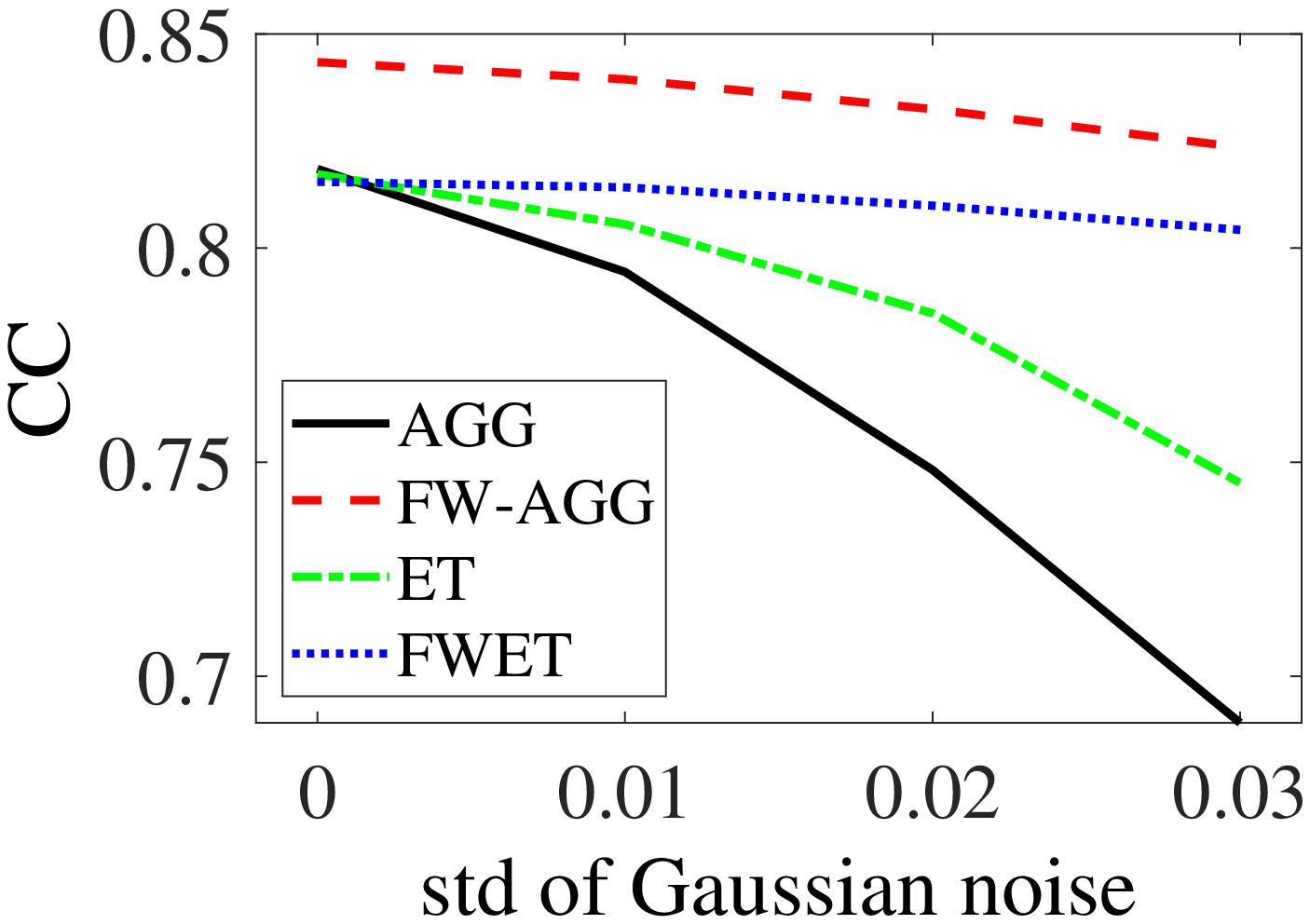}}
	\caption{(a) RMSE and (b) CC when Gaussian noise was added to the learned parameters of different algorithms. The models were trained on Subjects~1,~3-15 and tested on Subject~2.}	\label{fig:sharpminima}
\end{figure}

As shown in Fig.~\ref{fig:sharpminima}, FW-AGG was more robust to the perturbations than AGG, and FWET was also more robust than ET. These observations demonstrated that FW led the model to flatter minima in the parameter space, which helped improve its generalization ability.

\begin{figure}[htpb]	\centering
	\subfigure[]{\includegraphics[width=0.7\linewidth,clip]{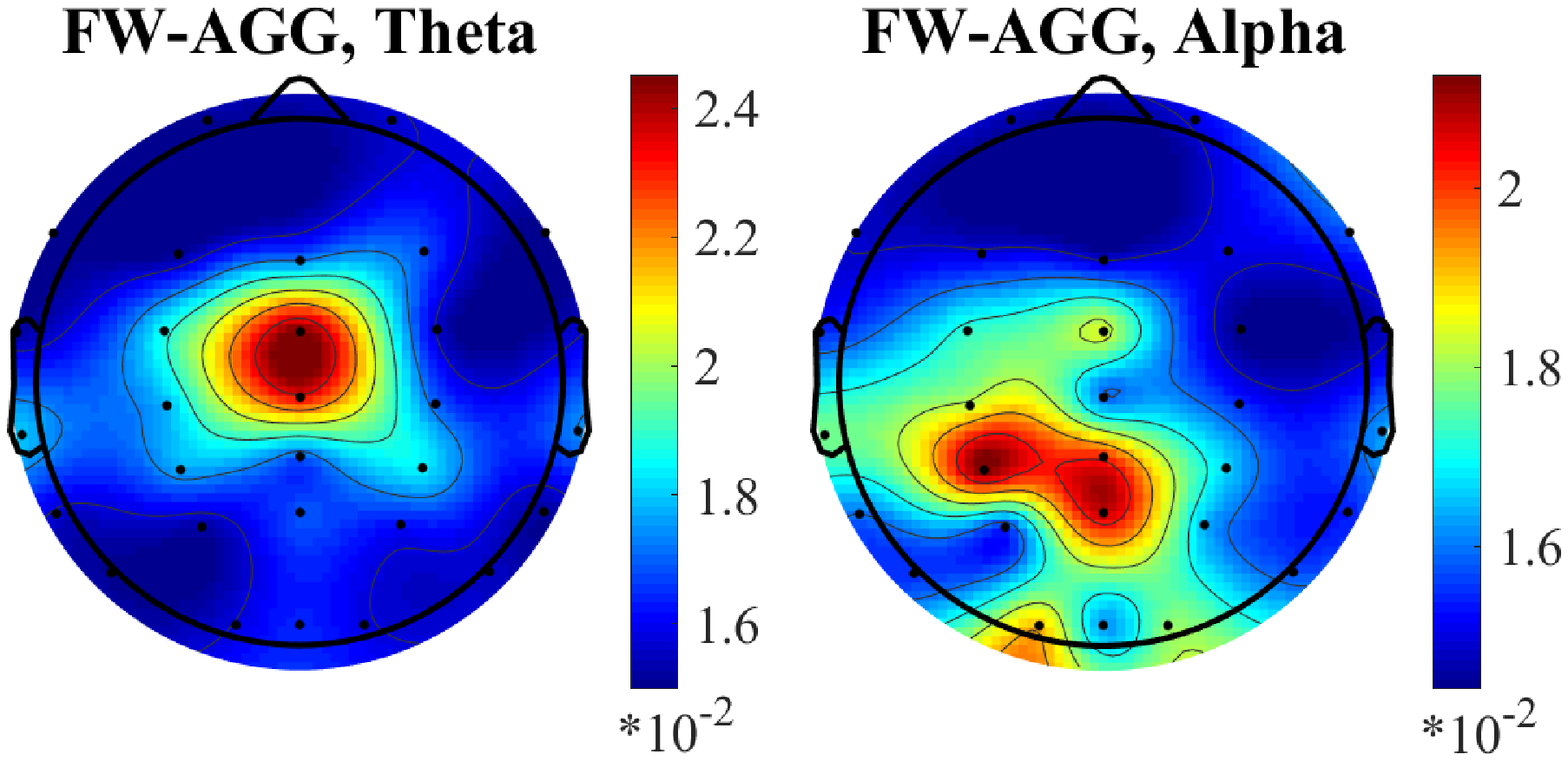}\label{fig:topo1}}
	\subfigure[]{\includegraphics[width=0.7\linewidth,clip]{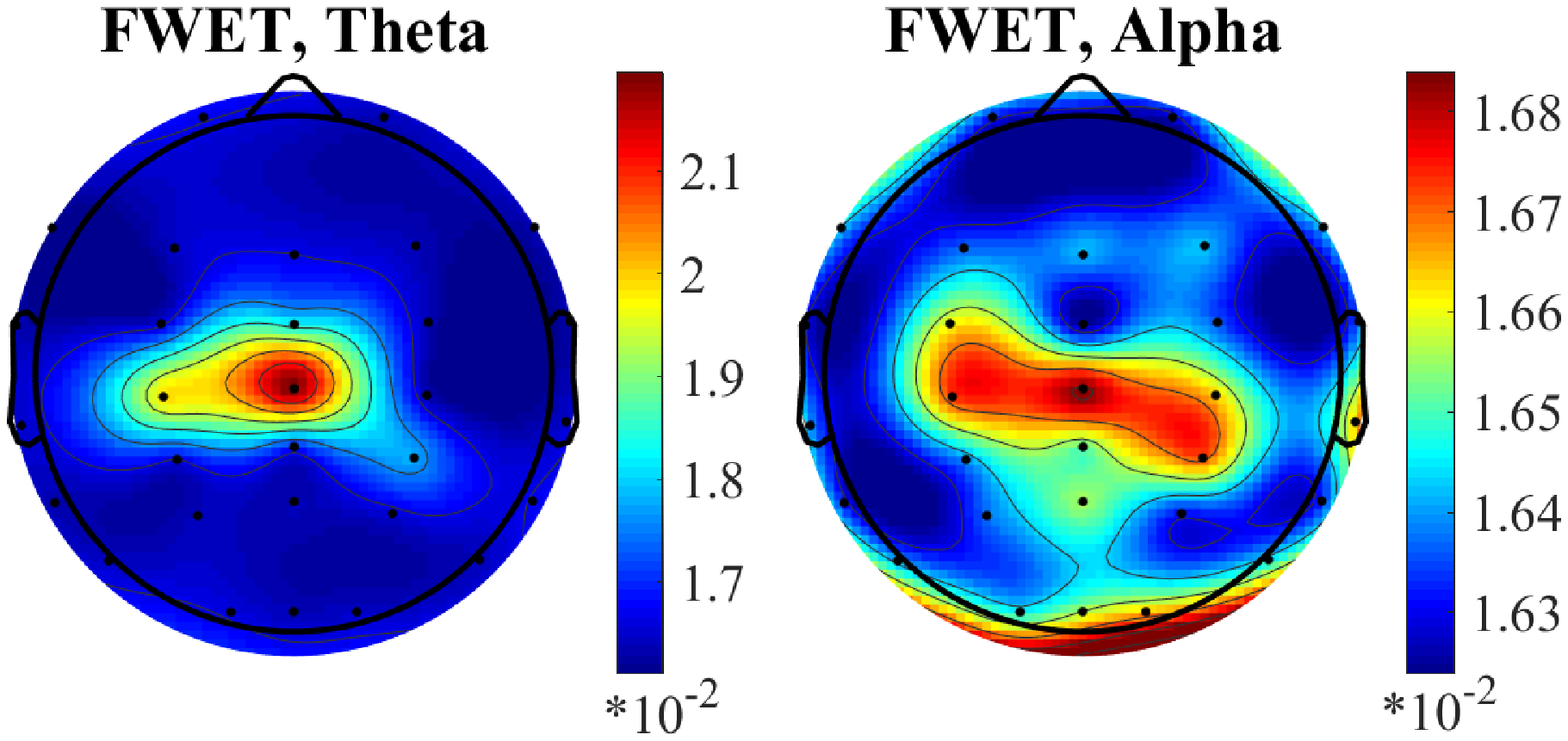}\label{fig:topo2}}
	\caption{EEG channel importance in theta and alpha bands, converted from $softmax(\mathbf{w})$ in (a) FW-AGG and (b) FWET.}	\label{fig:topo12}
\end{figure}

We also visualize the importance of different regions in each power band, determined by $\mathbf{w}$ in FW.  Fig.~\ref{fig:topo1} shows the topoplots of $\mathbf{w}$ in theta and alpha bands after the $softmax$ function in FW-AGG, when the last 14 subjects were used in training. For the theta band, the central brain region had the maximum weights, i.e., it contributed the most to drowsiness estimation. For the alpha band in FW-AGG, both the central and the occipital brain regions contributed more to drowsiness estimation than other regions. These were partially consistent with the findings in \cite{zhao2012electroencephalogram}, where Zhao \emph{et al.} studied mental fatigue in 90-minute continuous simulated driving, and found that the frontal, \emph{central} and occipital regions in the theta band, and the \emph{central}, parietal, \emph{occipital} and temporal regions in the alpha band, all had significant difference at the beginning and the end of the driving.

Fig.~\ref{fig:topo2} shows the topoplots of $\mathbf{w}$ in theta and alpha bands after the $softmax$ function in FWET, when the last 14 subjects were used in training. We can observe roughly the same patterns as in Fig.~\ref{fig:topo1} for FW-AGG. However, note that the magnitude ranges in Fig.~\ref{fig:topo2} were much smaller than those in Fig.~\ref{fig:topo1}, i.e., $\mathbf{w}$ in FWET had smaller variance than that in FW-AGG.

\begin{figure}[htpb]	\centering
	\subfigure[]{\includegraphics[width=0.7\linewidth,clip]{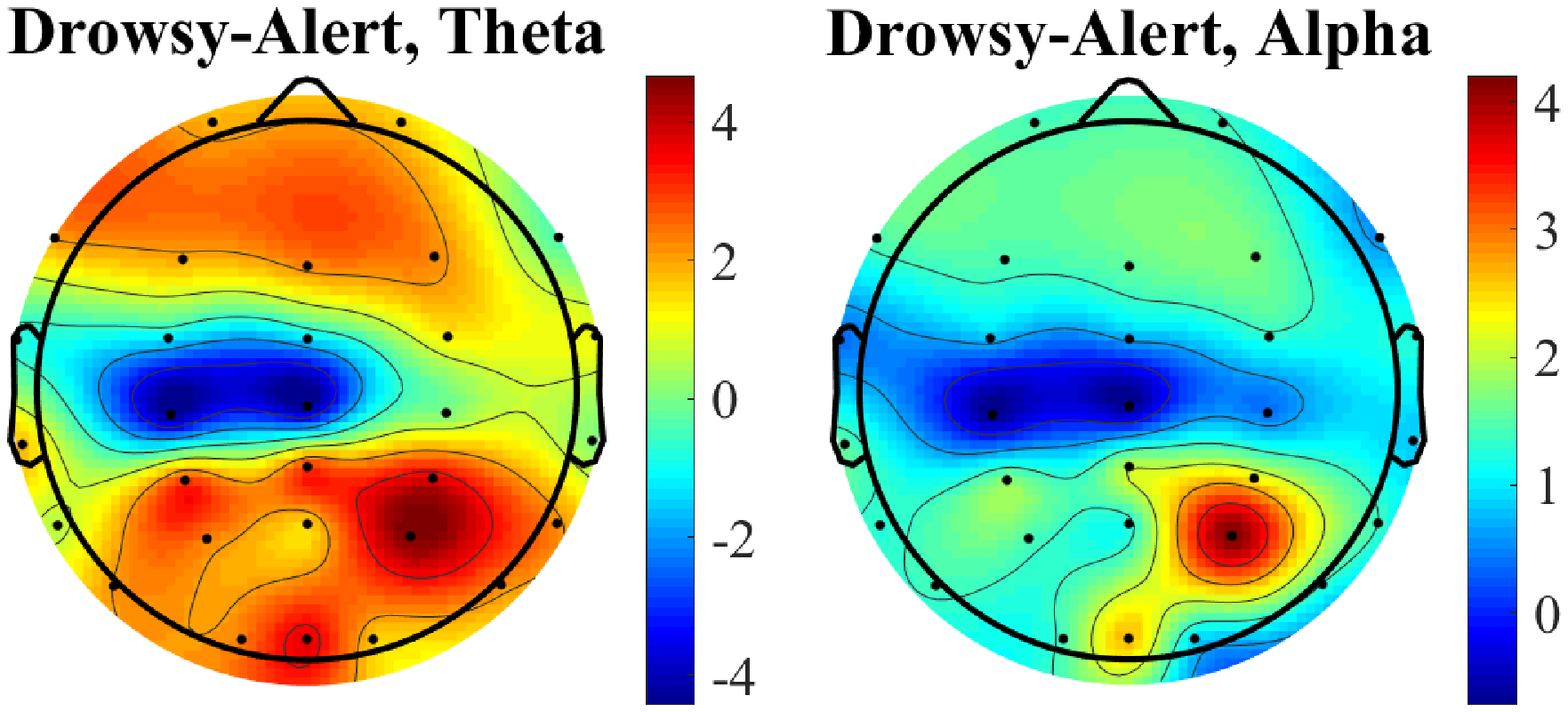}\label{fig:topo3}}
	\subfigure[]{\includegraphics[width=0.7\linewidth,clip]{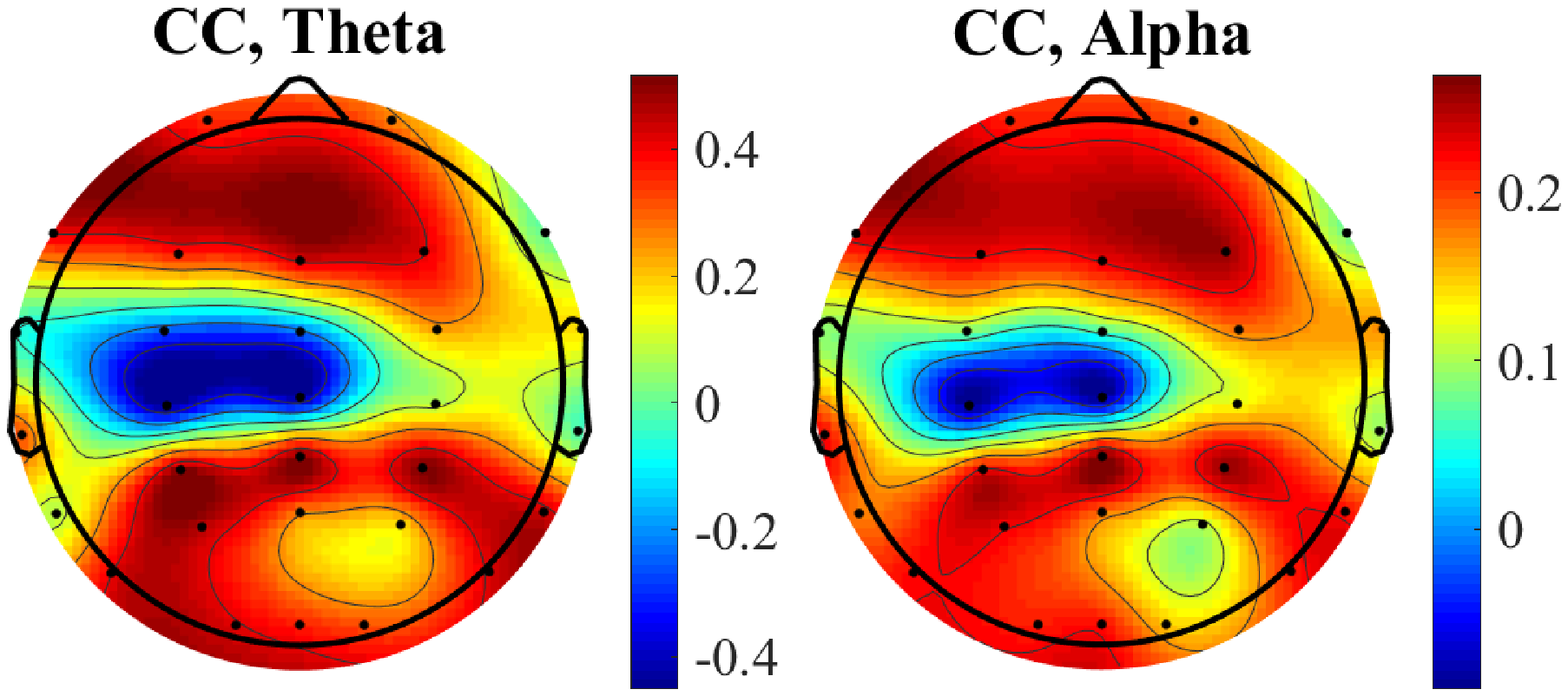}\label{fig:topo4}}
	\caption{(a) the differences between the topoplots of the drowsy and alert states. (b) the Pearson correlation coefficient between each PSD feature and the DI.}	\label{fig:topo34}
\end{figure}

We also computed the average PSD values for alert and drowsy states over the 15 subjects. We considered the subject be alert (drowsy) when his/her DI was lower (higher) than the 15 (85) percentile of the DIs over the entire session. Fig.~\ref{fig:topo3} shows the differences between the topoplots of the drowsy and alert states, and Fig.~\ref{fig:topo4} the Pearson correlation coefficient between each PSD feature and the DI. Interestingly, Figs.~\ref{fig:topo2} and \ref{fig:topo4} are not similar, i.e., though the channel weights $\mathbf{w}$ helped improve the drowsiness estimation performance, they were different from the correlation coefficients between the corresponding features and the DI.

Finally, although FW looks similar to the attention mechanism~\cite{vaswani2017attention}, which is being widely used in computer vision and natural language processing, they are different. The attention mechanism assigns dynamic weights to the neighboring locations, which change as the input varies. FW uses a fixed weight for each EEG channel, as the contributions of different brain regions usually do not change much in the same mental task.

\subsection{Effects of ET}\label{sec:etres}

This subsection first presents two experiments to understand how ET helped extract more generalizable features from different subjects, and then studies the effect of adding more regularization terms in ET and FWET.

\begin{figure}[htpb] 	\centering
	\subfigure[]{\includegraphics[width=0.9\linewidth,clip]{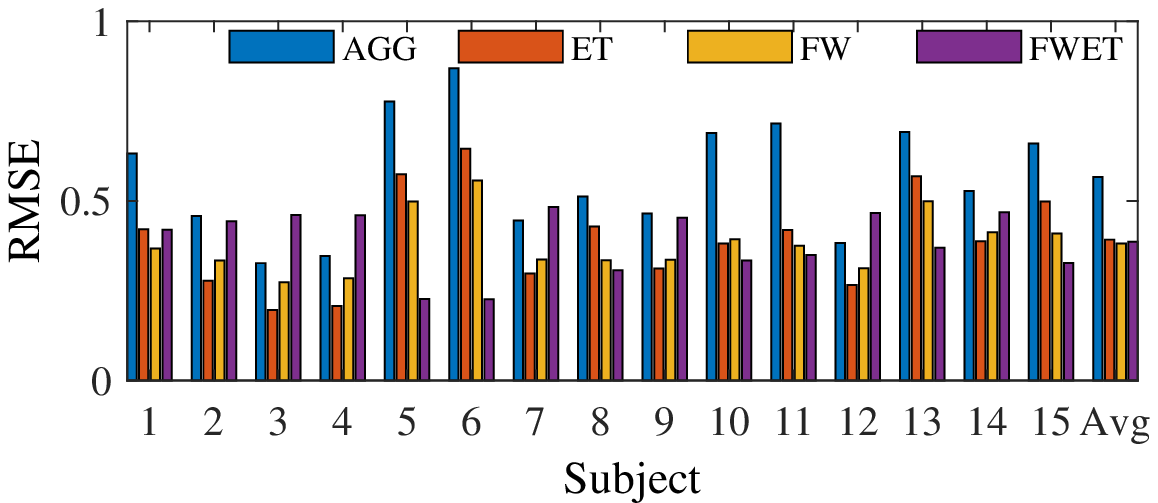}}
	\subfigure[]{\includegraphics[width=0.9\linewidth,clip]{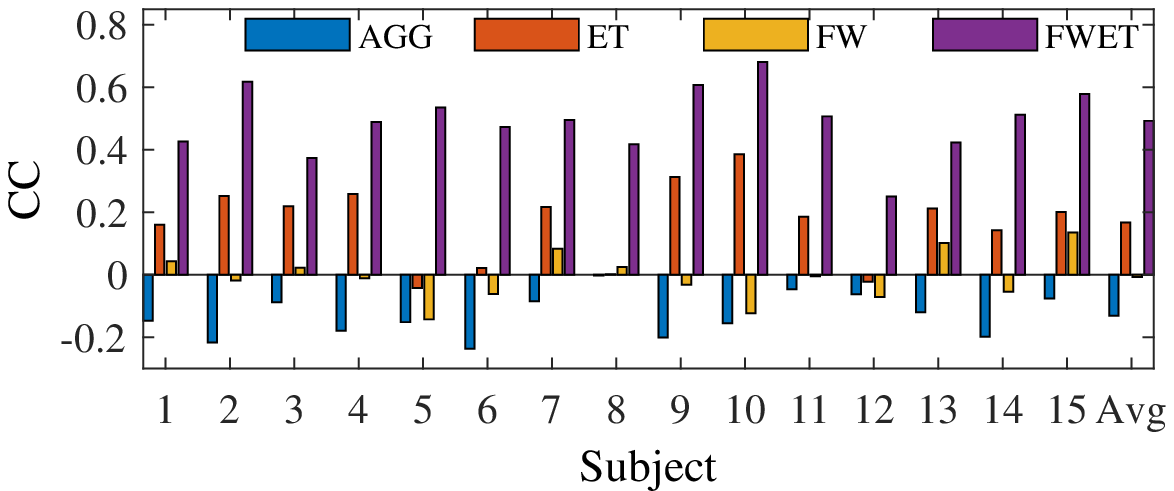}}
	\caption{Average RMSEs (a) and CCs (b) when Subject~$j$'s regression network $F_{\bm{\psi}_j}$ was applied to data from Subject~$s$ ($s\neq j$). The feature transformation was $F_{\bm{\theta}}(\bm{x})$ for both AGG and ET. The feature transformation was $F_{\bm{\theta}}(\hat{\mathbf{w}}\circ \bm{x})$ for both FW-AGG and FWET.} 	\label{fig:mismatchesET}
\end{figure}

We used data from all 15 subjects to train AGG, ET, FW and FWET, which had different feature extractor $F_{\bm{\theta}}$. To compare these two $F_{\bm{\theta}}$, we input Subject~$s$'s data to each $F_{\bm{\theta}}$, and used Subject~$j$'s ($j\neq s$) regressor $F_{\bm{\psi}_j}$ (which was trained on data from Subject~$j$ only) for regression. For AGG and ET, the final regression model was $F_{\bm{\psi}_j}(F_{\bm{\theta}}(\mathbf{x}^s))$. For FW-AGG and FWET, the final regression model was $F_{\bm{\psi}_j}(F_{\bm{\theta}}(\hat{\mathbf{w}}\circ \mathbf{x}^s))$. We tried all $j\neq s$ for each $s$, i.e., 14 different $F_{\bm{\psi}_j}$, and computed the average performance for each $s$. The smaller (larger) the RMSE (CC) is, the better the generalization performance is. The results are shown in Fig.~\ref{fig:mismatchesET}, where the subject index means that subject's data were used as input (Subject~$s$ in the above description). ET always achieved a smaller RMSE and a larger CC, suggesting that ET extracted more generalizable features. FWET had comparable RMSEs as FW-AGG, but generally larger CCs than FW-AGG,  suggesting again that ET extracted more generalizable features.

Three different regularizations were used in ET in \cite{li2019episodic} for classification problems:
\begin{enumerate}
\item \emph{epif} (short for \emph{episodic feature}), which requires the trained \emph{feature extractor} to work well with all domain-specific \emph{classifiers}.
\item \emph{epic} (short for \emph{episodic classifier}), which requires the trained \emph{classifier} to work well with all domain-specific \emph{feature extractors}.
\item \emph{epir} (short for \emph{episodic random}), which requires the \emph{feature extractor} to work well with a randomly initialized \emph{classifier} (representing a completely new domain).
\end{enumerate}

We only adopted \emph{epif} in our ET, because it was much easier and faster to optimize. The average training time per iteration, when different regularization terms were used in ET and FWET, are shown in Table~\ref{tab:et_simp_comp}. Intuitively, the computational cost increased when more regularization terms were used.

Table~\ref{tab:et_simp_comp} also shows the RMSEs and CCs when more regularizations were used. The weights for the three regularization terms were all set to $0.1$. For both ET and FWET, using \emph{epif} only achieved comparable performance with models using more regularization terms, and sometimes even slightly better. For the same type of regularization, FWET always outperformed ET, suggesting again the benefit of FW.

\begin{table}[htpb]\centering \setlength{\tabcolsep}{5mm}
\caption{Average RMSEs, CCs and training time (s) when different regularization terms were used in ET and FWET. }
\begin{tabular}{l|cc|c}
\hline
                               & RMSE            & CC              & Time (s)          \\ \hline
ET (ET-epif)                   & 0.2621          & 0.5434          & \textbf{0.5302}   \\
ET-epif-epic                   & \textbf{0.2577} & \textbf{0.5486} & 0.7050            \\
ET-epif-epic-epir              & 0.2682          & 0.5230          & 0.9235            \\ \hline
FWET (FWET-epif)               & \textbf{0.2332} & \textbf{0.5989} & \textbf{0.9651}   \\
FWET-epif-epic                 & 0.2398          & 0.5771          & 1.0530            \\
FWET-epif-epic-epir            & 0.2384          & 0.5795          & 1.2812            \\ \hline
\end{tabular} \label{tab:et_simp_comp}
\end{table}

In summary, we have shown that our proposed ET and FWET are efficient and effective, and their extracted features have comparable (or even slightly better) generalization performance with those with more regularization terms.

\subsection{Performance Gap between Validation and Test}

Early stopping on a validation set is frequently used in machine learning to reduce overfitting, and was also the case in this paper. However, the validation performance is usually more optimistic than the test performance. A model with stronger generalization ability should have a smaller performance gap between the validation performance and the test performance.

\begin{figure}[htpb] \centering
\includegraphics[width=.8\linewidth,clip]{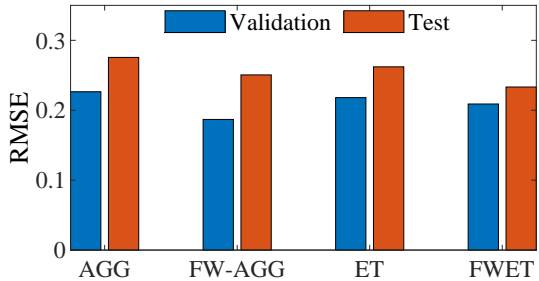}
	\caption{Validation and test RMSEs of the four AGG-based algorithms. A smaller gap between the validation RMSE and the test RMSE indicates better generalizability.} 	\label{fig:gap}
\end{figure}

Fig.~\ref{fig:gap} shows the validation and test RMSEs of the four AGG-based algorithms. Although AGG had the smallest validation RMSE, its test RMSE was the largest, i.e., the performance gap between the validation and test RMSEs were the largest, suggesting poor generalization ability. The validation-test RMSE gaps of FW-AGG, ET and FWET were considerably reduced. Particulary, FWET had the smallest gap, and the best test RMSE, suggesting its strong generalizability.

\subsection{Individualized $\tau_0$}\label{sec:adaptive}

$\tau_0=1$ in (\ref{eq:di}) was used in all above experiments. This is because we considered the most challenging case in brain-computer interfaces, i.e., we do not have any labeled data from the new subject. However, if we have some labeled data from the new subject, or some prior knowledge about the reaction time of the new subject, then it is possible to set $\tau_0$ individually. This subsection demonstrates the performance of FWET in this case.

\begin{figure}[htpb] \centering
\subfigure[]{\includegraphics[width=0.95\linewidth,clip]{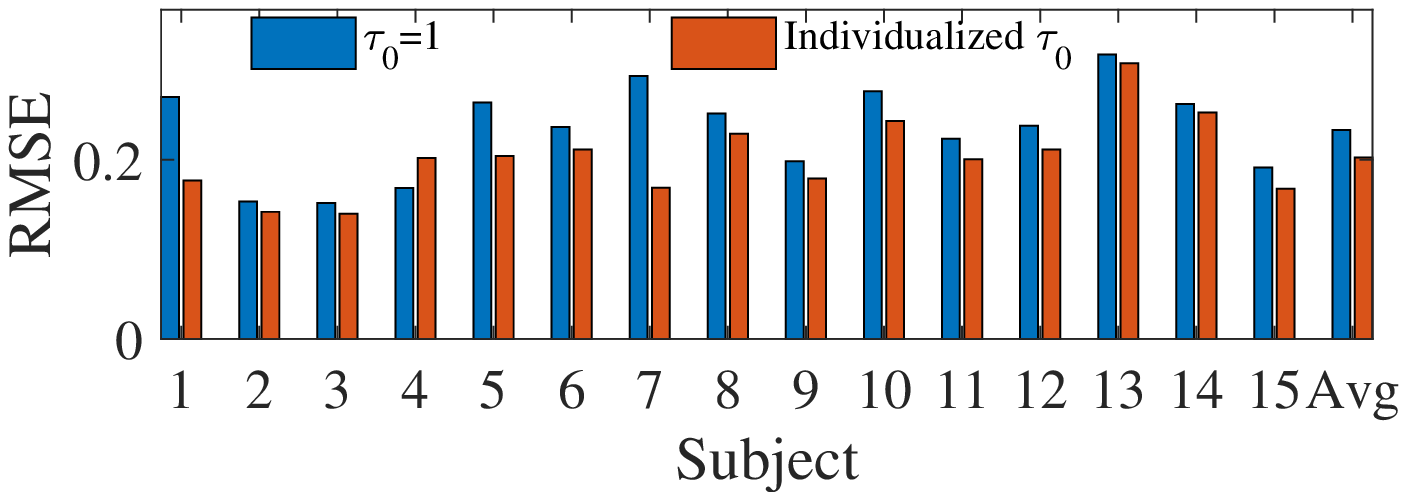}}
\subfigure[]{\includegraphics[width=0.95\linewidth,clip]{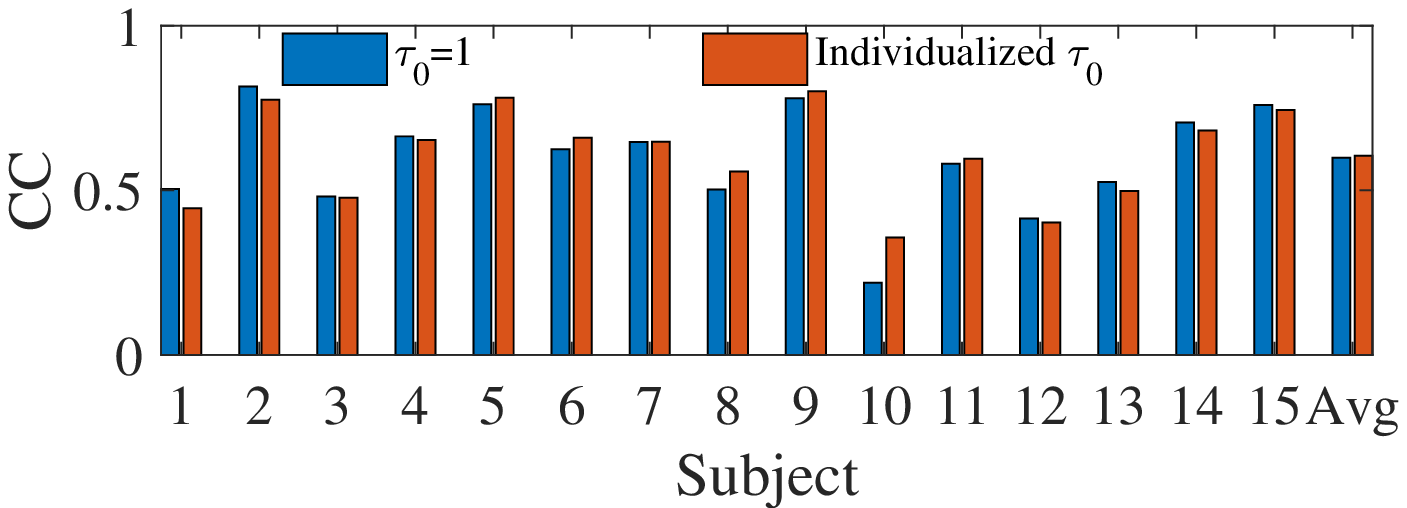}}
\caption{(a) RMSEs and (b) CCs in leave-one-subject-out cross-validation of FWET, using $\tau_0=1$ and individualized $\tau_0$.}	 \label{fig:bar_adaptive_tau}
\end{figure}

Following the practice in \cite{wei2015selective}, we set $\tau_0$ in (\ref{eq:di}) to be 5 percentile value of the reaction time of the corresponding subject, and repeated the experiments. The performances of FWET for constant and individualized $\tau_0$ are shown in Fig.~\ref{fig:bar_adaptive_tau}. Using individualized $\tau_0$ reduced the RMSE for almost every subject (except Subject~4), although the CCs were roughly the same. This demonstrates that more information about the new subject can generally improve the drowsiness estimation performance.


\subsection{Discussion} \label{sec:discussion}

This paper extends domain generalization, a concept mostly used in computer vision, to brain-computer interfaces. There are some important differences between these two application areas, which should be paid attention to in future research:
\begin{enumerate}
\item \emph{The number of source domains}. In computer vision applications, the number of domains is usually small, e.g., PACS\footnote{https://domaingeneralization.github.io/} has four domains, IXMAS\footnote{http://4drepository.inrialpes.fr/pages/home} has five domains, and MNIST\footnote{http://yann.lecun.com/exdb/mnist/} usually has seven rotated domains. Scalability is usually ignored in such applications. However, in brain-computer interfaces, more and more datasets with a large number of subjects are collected, and the scalability with respect to the number of domains can no longer be ignored. 

\item \emph{The variation of the label distribution in different domains}. Most existing domain generalization approaches only focus on learning a feature transformation $T$ that makes all source domains to have roughly the same marginal distribution $P(T(X))$, without considering the label distribution $P(Y)$. In EEG-based driver drowsiness estimation, the distribution of DIs varies significantly among different subjects. This makes generalization across different subjects difficult in brain-computer interfaces. 
\end{enumerate}

\section{Conclusion}\label{sec:conclusion}

EEG-based driver drowsiness estimation could be very important in improving driving safety. Unfortunately, individual differences among different drivers make it very difficult to design a generic estimation algorithm, whose parameters are fixed and optimal for all subjects. Usually some subject-specific calibration data are needed to tune the model parameters before applying it to a new subject, which is very inconvenient and not user-friendly. Many approaches have been proposed to reduce this calibration effort, but few can completely eliminate it. This paper has proposed an FWET approach to completely eliminate the calibration requirement. It integrates two techniques: FW to learn the importance of different features, and ET for domain generalization. Experiments demonstrated that both FW and ET are effective, and their integration can further improve the generalization performance. FWET does not need any labelled or unlabelled calibration data from the new subject at all, and hence could be very useful in plug-and-play brain-computer interfaces. Our future research will apply FWET to more such applications.


\end{document}